\documentclass[aip,pof,reprint,groupedaddress,superscriptaddress]{revtex4-1}

\usepackage{graphicx}
\usepackage{hyperref}
\usepackage{subfigure}
\usepackage{bm}
\usepackage{color}
\usepackage[normalem]{ulem}

\newcommand{\oca}{Laboratoire Lagrange, Universit\'e de Nice-Sophia
  Antipolis, CNRS, Observatoire de la C\^ote d'Azur, Nice, France}
\newcommand{\mpids}{Max Planck Institute for Dynamics and
  Self-Organization, G\"ottingen, Germany}

\begin{document}

\title{Turbulence attenuation by large neutrally buoyant particles}

\author{M.\ Cisse} \affiliation{\oca} \author{E.-W.\ Saw}
\affiliation{\oca} \author{M.\ Gibert} \affiliation{Universit\'{e}
  Grenoble Alpes, Inst.\ NEEL, F-38042 \hbox{Grenoble, France}}
\affiliation{CNRS, Inst.\ NEEL, F-38042
  \hbox{Grenoble, France}}\author{E.\ Bodenschatz}
\affiliation{\mpids} \affiliation{Institute for Nonlinear Dynamics,
  University of \hbox{G\"ottingen, G\"ottingen, Germany}}
\affiliation{Laboratory of Atomic and Solid-State Physics and
  Sibley~School~of~Mechanical and Aerospace Engineering, Cornell
  University, USA} \author{J.\ Bec} \affiliation{\oca}

\begin{abstract}
  Turbulence modulation by inertial-range-size, neutrally-buoyant
  particles is investigated experimentally in a von K\'arm\'an
  flow. Increasing the particle volume fraction $\Phi_\mathrm{v}$,
  maintaining constant impellers Reynolds number attenuates the fluid
  turbulence. The inertial-range energy transfer rate decreases as
  $\propto\Phi_\mathrm{v}^{2/3}$, suggesting that only particles
  located on a surface affect the flow.  Small-scale turbulent
  properties, such as structure functions or acceleration
  distribution, are unchanged. Finally, measurements hint at the
  existence of a transition between two different regimes occurring
  when the average distance between large particles is of the order of
  the thickness of their boundary layers.
\end{abstract}

% \pacs{52.30.-q, 52.65.-y, 52.30.Cv}

\maketitle

Introducing impurities in a developed turbulent flow has drastic
effects on the mechanisms of energy transfer and dissipation.  A
minute amount of polymer additives causes for instance drag
reduction.\cite{WM08} In such visco-elastic fluids, the coupling
between the local flow and the polymers stretching can be modeled in
order to quantify the exchanges between kinetic and elastic energies
and to interpret turbulent drag reduction as a suppression of large
velocity gradients.\cite{PMP06} The mechanisms at play in turbulent
suspensions of finite-size spherical particles are much more intricate
and turbulence can either be enhanced or suppressed.\cite{BE10} In
wall flows, transition to turbulence can be either hindered or
facilitated depending on the size and volume fraction of
particles.\cite{MMG03,LAMC13} In the case of developed turbulence,
depending on the size of impurities, on their density ratio, and
volume fraction, both turbulent kinetic energy and dissipation rate
can either increase or decrease.\cite{LFE10,BBC+12} There is no thorough
understanding of turbulence modulation by finite-size particles. It is
only in the asymptotics of very small and very dilute particles that
one can satisfactorily model the energy budget between the fluid and
the dispersed phase.  Away from these limits, either finite-size
effects preclude an explicit formulation of the forces exerted on the
particles or large concentrations require the understanding of
possible collective effects.

Finite-size effects are usually accounted for in terms of Fax\'en
corrections to the Stokes equation governing point-particle
dynamics.\cite{Gat83,MR83} However, it was observed\cite{HB10} that
this approximation works only for particle whose diameters is below
$\approx4\eta$, where $\eta$ designates the turbulent Kolmogorov
dissipative scale. The finiteness of the particle Reynolds number
becomes important at larger sizes and explicit models would require
solving the full non-linear Navier--Stokes equation with the proper
boundary conditions at the particle surface. A first step in tackling
finite-size effects consisted in characterizing the statistics of
various dynamical quantities, such as velocity, acceleration, rotation
rate, for large neutrally buoyant particles in developed turbulent
flows.\cite{QBB+07,XB08,ZGB+11} One of the main findings is that
average particle dynamical properties can be predicted from usual
turbulent dimensional analysis at scales of the order of the particle
diameter. More recently, much effort has been devoted to the
characterization of the fluid flow in the vicinity of a single
particle.\cite{BV12,CHB13,KGBB13} In particular, it was found that it
is surrounded by a boundary layer whose average thickness is of the
order of its diameter. At larger distances, the influence of the
particle onto the carrier fluid is negligible. Also, kinetic energy
dissipation is strongly increased at the particle surface (in its
viscous boundary sublayer) and weakly diminished in its
wake.\cite{TE10,CHB13}

All these observations for neutrally spherical particles were obtained
in the limit where they are very dilute, so that individual
disturbances of the carrier never pile up nor interfere. At higher
concentrations, large particles interact together through either
collisions or perturbations of the carrier flow. This ``four-way
coupling'' occurs when particles are at distances less than the
thickness of their boundary layer. In principle, if the particles are
uniformly distributed in the flow with a volume fraction
$\Phi_\mathrm{v}\ll 1$, the effects of two-particles interactions are
expected to be $\propto \Phi_\mathrm{v}^2$, and are thus negligible
compared to single-particle contributions which are $\propto
\Phi_\mathrm{v}$. Several works indicate that such a naive picture
cannot be true. First, neutrally-buoyant particles seem to distribute
in a non uniform manner; they cluster near walls\cite{CL06,KCDU13} or
correlate with the large scales of the carrier flow.\cite{MZF+14} It
is thus probable that local particle density has strong fluctuations.
Second, large particles that approach very close to each other might
spend some time together. This is suggested for instance in the
simulations of Ref.~\onlinecite{TDP+04} and could be due to strong
dissipative mechanisms occurring when particles meet. This effect
could lead to the creation of particle clumps, as for instance
observed for heavy small-size elastic particles.\cite{BMR13} Such
collective phenomena, if exist, might be crucial in the understanding
of turbulence modulation by finite-size particles.

We present here experimental results on the influence of finite-size,
neutrally buoyant particles on a turbulent von K\'arm\'an flow. We
find that the most noticeable effects induced by the dispersed phase
is a continuous decrease of the turbulent kinetic energy and of the
average inertial-range transfer rate when increasing the particle
volume fraction while keeping constant the rotation speed of the
impellers. We also obtain evidence that the small-scale turbulent
properties of the fluid flow are unchanged by the presence of the
particles. In particular, our results show surprisingly that in the
presence of particles, second-order statistics of velocity increments
match those of unladen flows. Finally, our results support the
existence of a transition between two different regimes which occurs
when the average distance between large particles is of the order of
the thickness of their boundary layers.

%\bigskip 
%\noindent\emph{Experimental setup}
%\smallskip

We consider a von K\'arm\'an water flow maintained in a developed
turbulent state (with $R_\lambda \approx 300$) by two impellers of
diameter $d = 28\,$cm that counter-rotate at an approximately constant
torque.  Specifically, we adjust the power of the motors at the
beginning of each experiment in order to assert the same average
rotational frequency of the impellers for all measurements. The
enclosure has an octagonal cylindrical shape with dimensions are
$40\times 38\times 38\,$cm. The flow field is analyzed by tracking the
temporal evolution of fluorescent tracer particles
($D\approx 107\,\mu$m ) using three cameras (Phantom V10, manufactured
by Vision Research Inc., Wayne, USA) at 2900 fps. The measurement
volume of approximate diameter $8\,$cm in the center of the tank is
illuminated by a 100W laser beam. We use the weighted averaging
algorithm for two-dimensional tracer finding and a ``three-frames
minimum acceleration'' for tracking their three-dimensional
trajectories.\cite{OXB06} The finite-size particles are
super-absorbent polymer spheres whose optical index and mass density
match those of water. These particles have diameters of the order of
$D_\mathrm{p} \approx 90 \eta \approx L/9$, where $\eta$ and $L$ are
the dissipative and integral length scales of the flow without
particles.  Two thin grids situated at $10\,$cm from the impellers
prevent particles from colliding with them.  This setup has been
already used in Ref.~\onlinecite{K12,KGBB13} and more details can be
found therein. We study here the fluid flow characteristics varying
the volume fraction of large particles from $\Phi_\mathrm{v}=0$ to
$10\%$. Each time the particle load is changed, the power of the
impellers is adjusted to maintain constant their rotation frequency at
$f_\mathrm{imp} = 0.79\,\mathrm{Hz}\ (\pm 1.5\%)$, and thus the
associated Reynolds number at
$Re = d^2\,f_\mathrm{imp}/\nu\approx62\,000$. The different setups are
summarized in Table~\ref{tab:parameters}.

\begin{table}[!ht] 
\centering
\begin{tabular}{|c||c|c|c|c|c|c|c|c|}
\hline
%$N_{p}$ & 0 & 50 & 124 & 300 & 600 & 1500 & 4500 \\
%\hline
$\Phi_\mathrm{v}$ & 0 &0.1\% &0.3\% &0.8\% & 2\% &4\%
&10\% \\ \hline
$f_{\rm imp}$ (Hz) &0.79 & 0.79 & 0.79 & 0.79 & 0.80 & 0.80 & 0.80 \\
\hline
$u_\mathrm{rms}$  (m\,s$^{-1}$) & 0.0796  &  0.0797 & 0.0771 & 0.0752 & 0.0766 & 0.0578 & 0.0466\\
\hline
%\bf $a_\mathrm{rms} \ (m\,s^{-2})$ & 1.3632 &1.3395 & 1.3061 & 1.2489 & 1.2901 &1.0573 & 0.9584\\
%\hline
$\varepsilon$ (m$^{2}$\,s$^{-3}$) & 0.0058 &0.0055 &0.0053  &0.0048 &0.0051  &0.0032 & 0.0012 \\
\hline
$\eta$ ($\mu$\,m) & 115  &  117 & 118 & 121 & 119& 133 & 183 \\
\hline
$\tau_\eta$ (m\,s) & 13&  14&  14 &  15 & 14 &  18 &  33  \\
\hline
$R_{\lambda}$ & 321 &335& 319 & 319 & 316 & 228 & 143\\
\hline
\end{tabular}
\caption{\label{tab:parameters} Characteristics of the various experiments. $\Phi_\mathrm{v}$
  volume fraction of the large particles; $f_{\rm imp}$ average
  frequency of the impellers; $u_\mathrm{rms}$ root-mean squared
  velocity (averaged over components); $\varepsilon$ average kinetic energy transfer rate
  (measured from second-order structure function; see text);
  $\eta=\nu^{3/4}/\varepsilon^{1/4}$ Kolmogorov dissipative scale; $\tau_\eta=\nu^{1/2}/\varepsilon^{1/2}$ associated
  turnover time; $R_{\lambda}$ Taylor-microscale Reynolds number.}
\end{table}

\begin{figure}[ht!]
  \centerline{
    \includegraphics[height=0.38\textwidth]{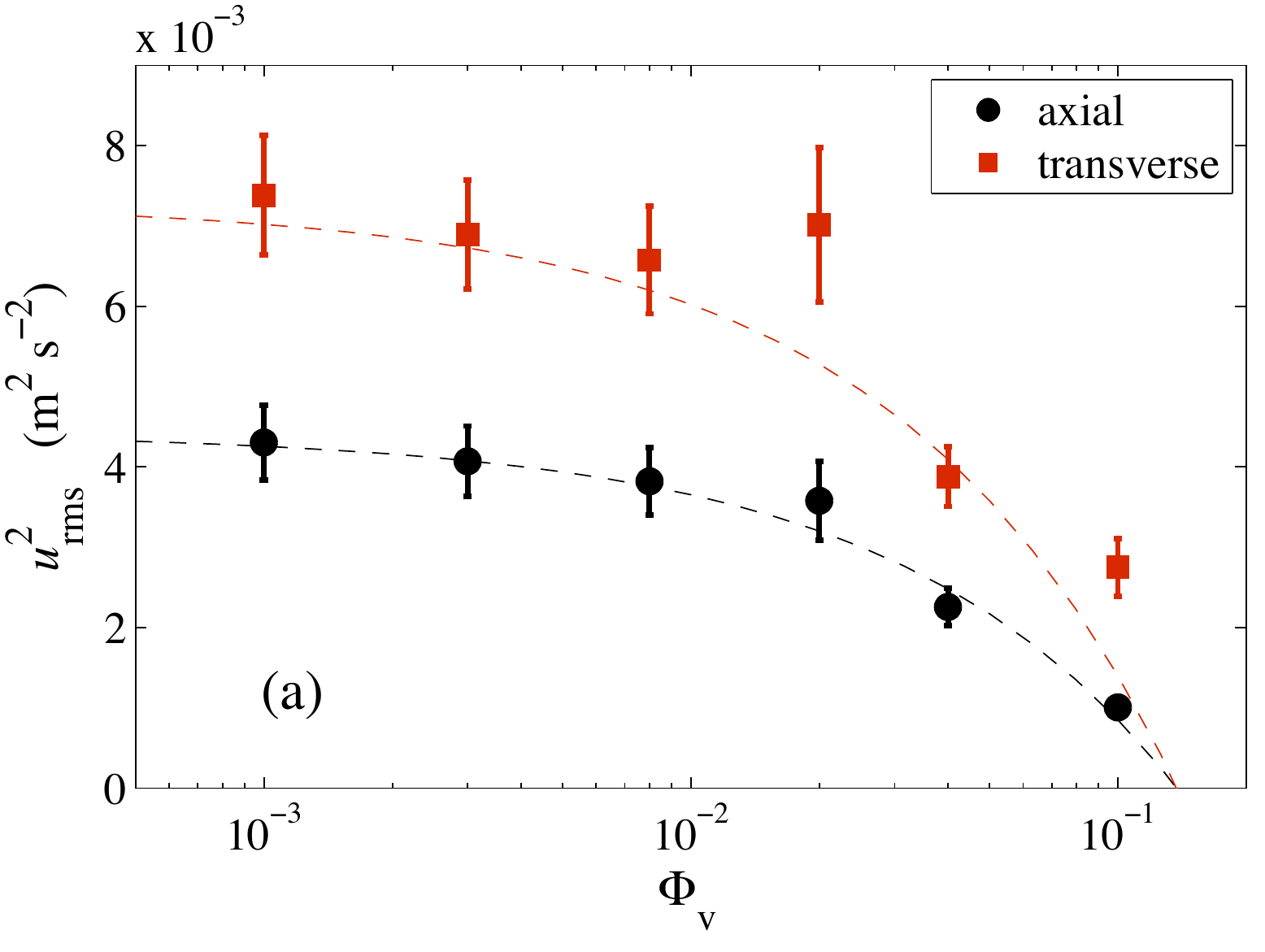}
    \hfill
    \includegraphics[height=0.38\textwidth]{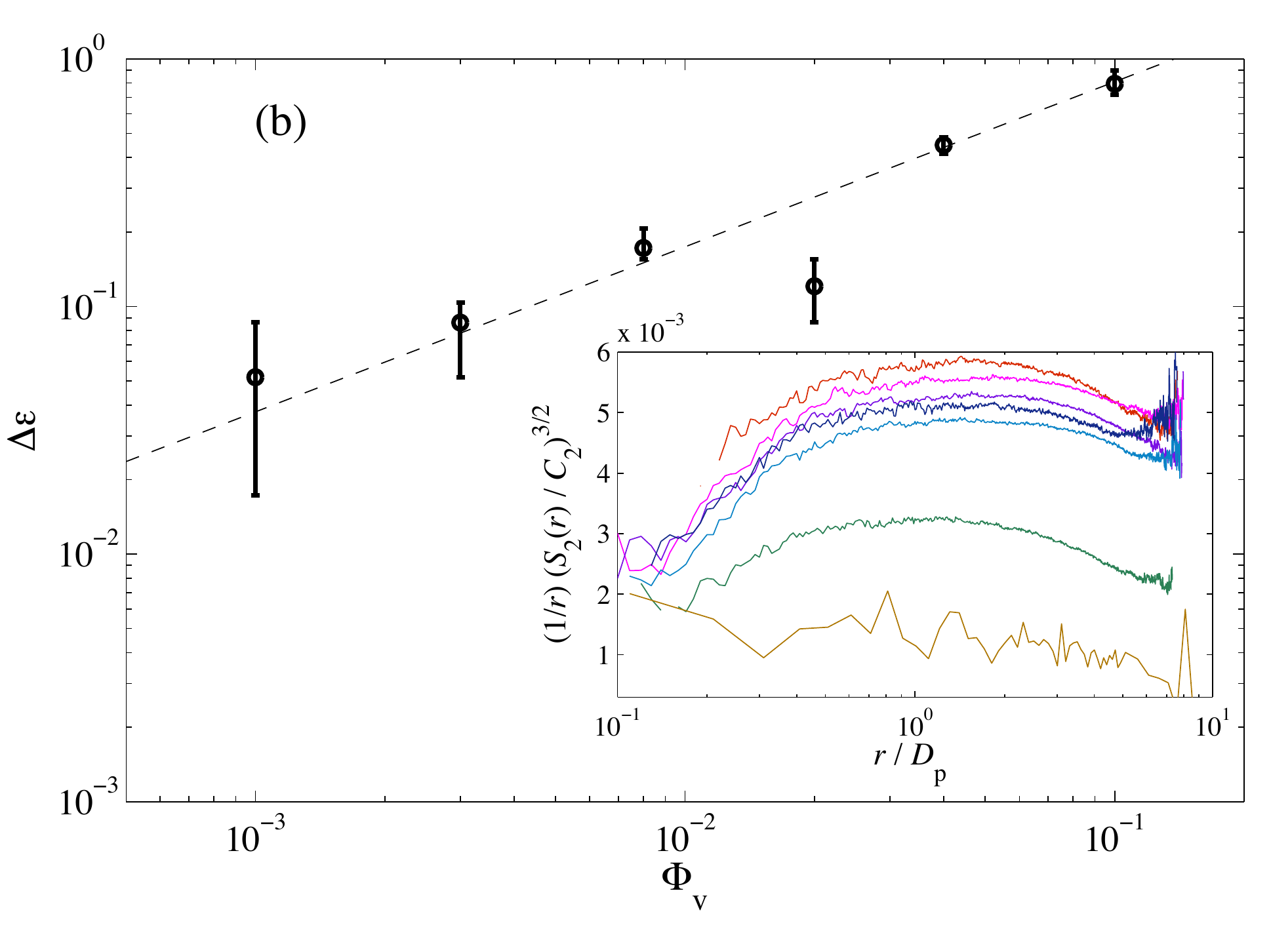}
  }
  \vspace{-10pt}
  \caption{\label{fig:u2_epsilon_fn_phi} (a) Variance of the turbulent
    velocity fluctuations
    $u_\mathrm{rms}^2 = \langle (u_i -\langle u_i\rangle)^2\rangle$ as
    a function of the particle volume fraction; The dashed lines are
    $u_\mathrm{rms}^2(\Phi_\mathrm{v}) =
    u_\mathrm{rms}^2(0)(1-3.75\,\Phi_\mathrm{v}^{2/3})$.
    (b) Discrepancy of the average inertial-range kinetic energy
    transfer rate
    $\Delta\varepsilon =
    1-\varepsilon(\Phi_\mathrm{v})/\varepsilon(0)$
    as a function of the particle volume fraction $\Phi_\mathrm{v}$;
    the dashed line is
    $\Delta\varepsilon = 3.75\,\Phi_\mathrm{v}^{2/3}$.  The rate
    $\varepsilon$ is estimated from the inertial-range average value
    of the compensated second-order longitudinal velocity structure
    functions $S_2(r)/(C_2 r^{2/3})$ using for the Kolmogorov constant
    $C_2=2.1$. \emph{Inset:} Dissipation rate
      obtained from the second-order Eulerian longitudinal structure
    functions for the various volume fractions (the upper curve
      corresponds to the case with no particles and $\Phi_\mathrm{v}$
      increases from top to bottom).}
\end{figure}

We start by considering the effect of the large particles on the
intensity of turbulent fluctuations. The Eulerian mean velocity
profile $\langle \bm u(\bm x)\rangle$ is obtained for the different
values of $\Phi_\mathrm{v}$ by binning positions and averaging
velocities over all movies. We then define velocity fluctuations as
instantaneous deviations of the tracers velocity from this average
field.  Figure~\ref{fig:u2_epsilon_fn_phi}(a) represents the variance
$u_\mathrm{rms}^2$ of the axial (along the axis of symmetry) and
transverse components of these turbulent velocity fluctuations as a
function of the volume fraction of large particles. One clearly
observes that the turbulent kinetic energy decreases when the particle
load increases, as already observed in a different turbulent
  flow.\cite{BBC+12} Data show anisotropy, which persists for all
$\Phi_\mathrm{v}$: the root-mean square values of axial velocity
components are around $60$-$70\%$ of the transverse ones. Measurements
are in good agreement with
$u_\mathrm{rms}^2(\Phi_\mathrm{v}) \approx
u_\mathrm{rms}^2(0)(1-3.75\,\Phi_\mathrm{v}^{2/3})$.
For transverse components, one point, corresponding to
$\Phi_\mathrm{v}\approx 2\%$ clearly deviates from this form. As we
will see later, this could be due to difficulties in estimating the
mean flow in that case. The fitting form predicts that
$u_\mathrm{rms}=0$ for $\Phi_\mathrm{v}\gtrsim 14\%$. This could
correspond to a critical volume fraction above which turbulence is
completely extinguished by the particles.

The inertial-range kinetic energy transfer rate $\varepsilon$ follows
a similar trend.  Figure~\ref{fig:u2_epsilon_fn_phi}(b) shows that the
discrepancy
$\Delta\varepsilon = 1-\varepsilon(\Phi_\mathrm{v})/\varepsilon(0)$ is
proportional to $\Phi_\mathrm{v}^{2/3}$, like for the turbulent
kinetic energy.  Even at small values of the volume fraction, the
attenuation of turbulence does not show a linear behavior
$\propto\Phi_\mathrm{v}$ (proportional to the total number of
  particles) that is expected if the effect of particles consists in
the superposition of non-interacting individual perturbations.  An
attenuation $\propto \Phi_\mathrm{v}^{2/3}$ suggests that collective
effects and interactions between particles are at play.  Such a
scaling indicates that only a fraction of the particles have an
important impact on the flow.  The power two-third can be interpreted
as if these active particles are located on a surface rather than in
the full volume of the experiment. Several heuristic scenario can lead
to such a law. First, we can imagine that particles form a given
number of aggregates where all fluid flow perturbations occur in the
boundary layers of the outer edge. We did not detect such aggregates
in the observation volume but we cannot exclude that they are located
in regions of the flow that we do not monitor (for instance close to
the grids near the impellers). Recent work found that particles are
likely to cluster far from the center of the von K\'arm\'an
flow.\cite{MZF+14} A more likely scenario relies on a kind of
shielding effect where the kinetic energy is prevented from reaching
the center of the flow by the particles situated at the periphery of
the experiment.  A full discrimination between these possibilities
requires systematically scanning the entire experiment in order to
measure the inhomogeneities in both the particle distribution and the
fluid turbulence characteristics.

%\bigskip 
%\noindent\emph{Small-scale turbulent properties}
%\smallskip

\begin{figure}[ht!]
  \centerline{
    \includegraphics[height=0.38\textwidth]{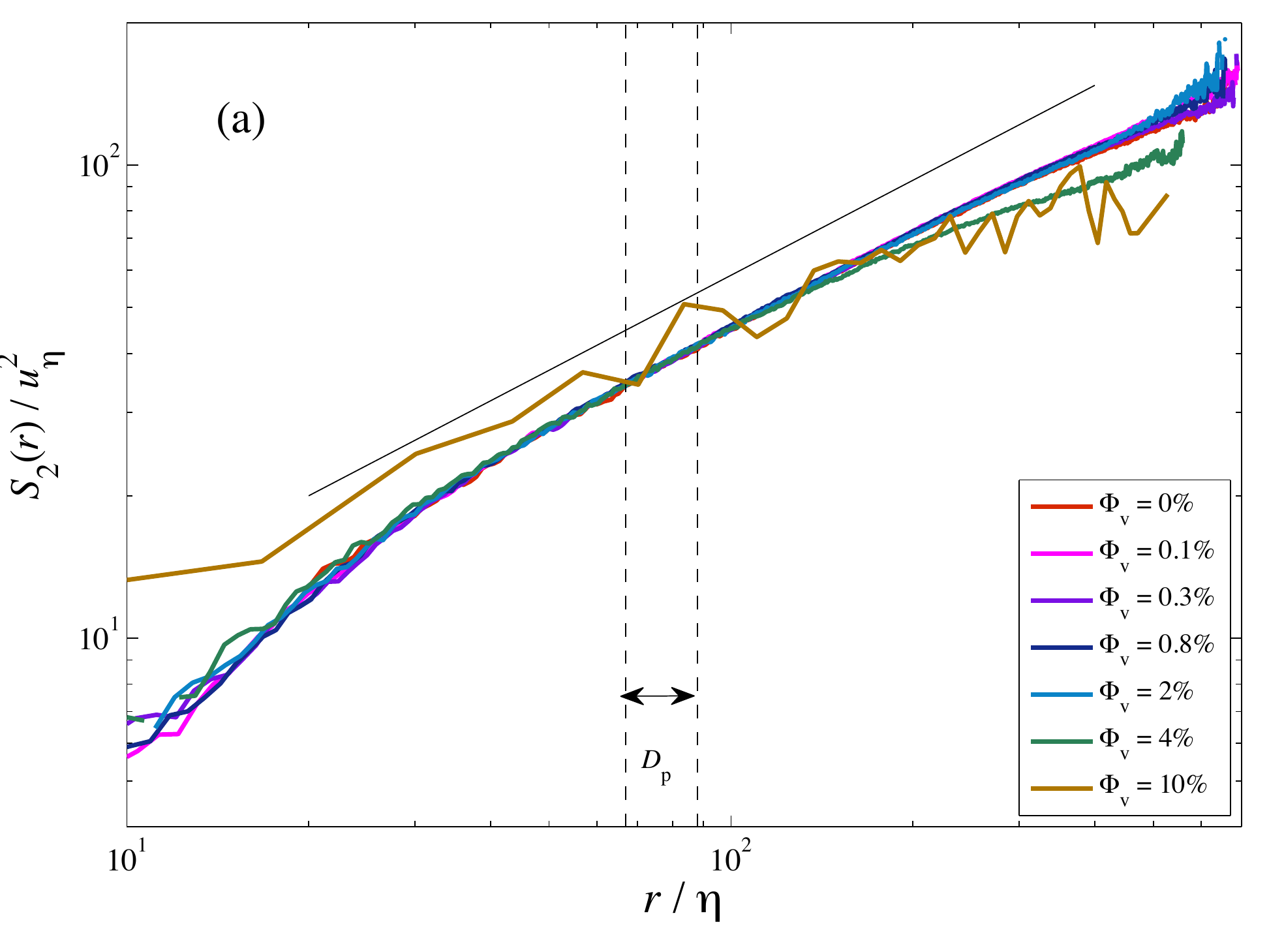}
  \hfill
  \includegraphics[height=0.38\textwidth]{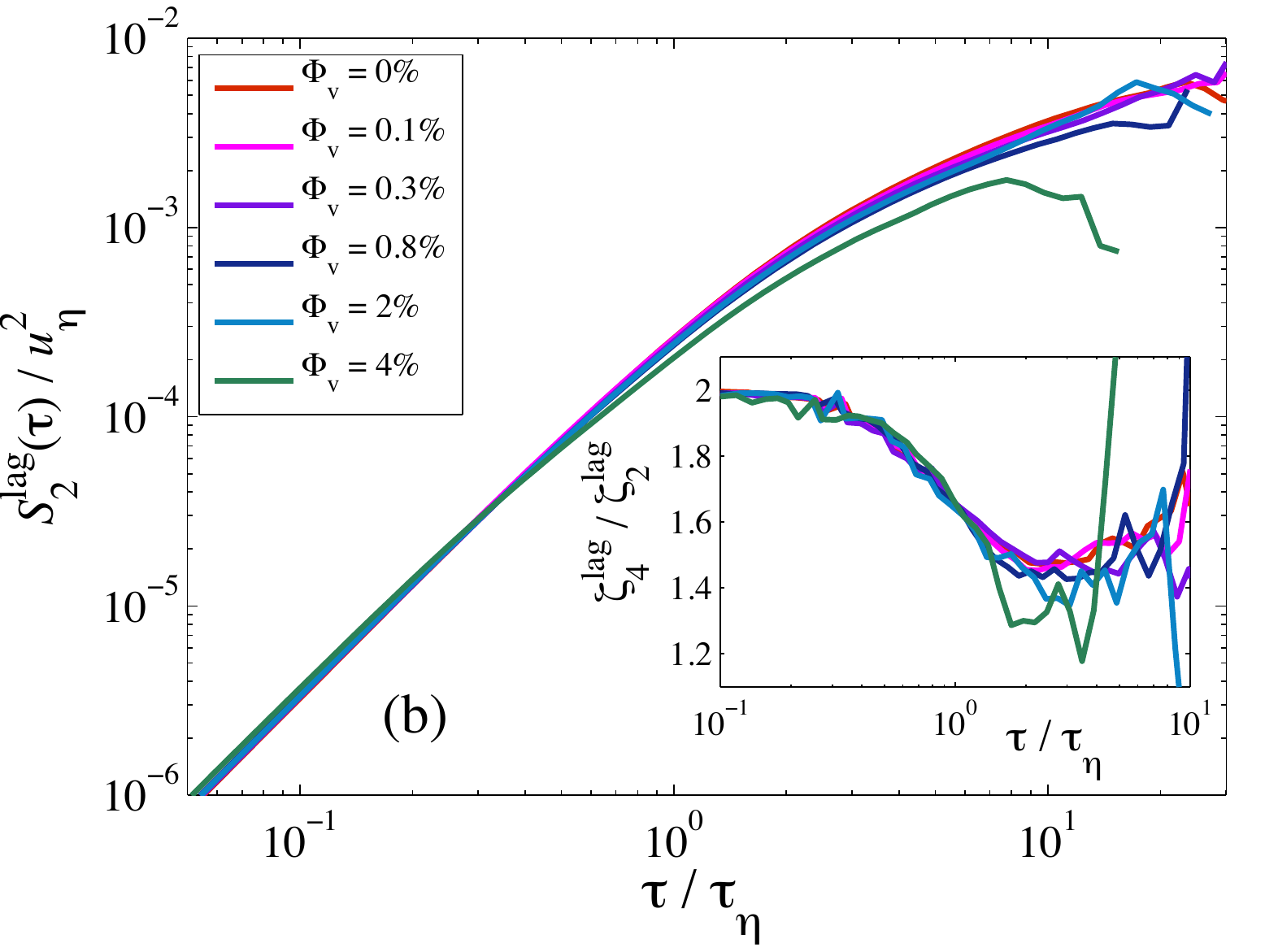} }
  \vspace{0pt}
  \caption{\label{fig:struct_fn} (a) Second-order Eulerian
    longitudinal structure functions $S_2(r)$ in dissipative-range
    units obtained for the various values of the volume fraction
    $\Phi_\mathrm{v}$; The solid line corresponds to a behavior
    $\propto r^{2/3}$; The two vertical dashed lines shows the range
    $54\lesssim D_\mathrm{p}/\eta\lesssim 88$ obtained from the
      maximal and minimal values of $\eta$. (b) Second-order
    Lagrangian structure function $S_2^\mathrm{lag}(\tau)$ in
    dissipative-scale units. \emph{Inset:} logarithmic derivative
    $(\mathrm{d}\log S_4^\mathrm{lag})/(\mathrm{d}\log
    S_2^\mathrm{lag})$
    of the fourth-order Lagrangian structure function with respect to
    the second-order.}
\end{figure}
We next move to small-scale turbulent properties. A striking result is
that they seem unaffected by the presence of the particles once scaled
with the corresponding transfer rate $\varepsilon$, large-scale
kinetic energy content $u_\mathrm{rms}^2$, and associated Reynolds
number $R_\lambda = u_\mathrm{rms}^2\sqrt{15/(\nu\varepsilon)}$.
Figure~\ref{fig:struct_fn}(a) shows the second-order longitudinal
Eulerian structure function of the turbulent velocity fluctuations for
the various particle loading considered. All curves collapse in the
inertial and sub-inertial ranges after they are represented in
dissipative-scale units. In particular there are no noticeable effects
(up to noise) at separations of the order of the particles
diameter for $\Phi_\mathrm{v}$ up to $10\%$.  These results seem to
contradict the kinetic energy spectra obtained in numerical
simulations\cite{TDP+04,YDCM10} that display an enhancement of the
energy contained at scales smaller than the particle size and a
depletion at larger scales.  These spectra are obtained from a Fourier
transform of the full domain encompassing both the fluid and the
particles. There are still questions whether such effects on the
spectrum are inherent to the perturbed fluid flow or come from the
particle velocity field. In our case, the use of Lagrangian tracers to
perform Eulerian measurements isolate the fluid statistics from such
pitfalls and suggest that there is no intrinsic modifications of the
distribution of turbulent velocity increments by the large particles.

Statistics along tracer trajectories seem also rather insensitive to
the presence of large particles.  Figure~\ref{fig:struct_fn}(b) shows
measurements of the Lagrangian structure functions $S^\mathrm{lag}_p =
\langle [u_i(t+\tau)-u_i(t)]^p \rangle$, where $u_i$ are components of
the tracer velocity and the average is over the ensemble of
trajectories. One observes that for $p=2$, the collapse is not as good
as for Eulerian statistics. However, Lagrangian structure functions
are known to display very intermittent properties and the observed
discrepancies are comparable to those obtained when comparing unladen
flows at different Reynolds numbers.\cite{BBC+08} We see from
Tab.~\ref{tab:parameters} that our Reynolds number decreases when the
particle volume fraction increases: $R_\lambda$ varies by more than a
factor two when $\Phi_v$ goes from $0$ to $10\%$. This lack of
universality is not present when considering only the scaling
properties through the logarithmic derivatives of the Lagrangian
structure functions.  For that reason, we follow
Ref.~\onlinecite{BBC+08} by showing in the inset of
Fig.~\ref{fig:struct_fn}(b) the quantity $(\mathrm{d}\log
S_4^\mathrm{lag})/(\mathrm{d}\log S_2^\mathrm{lag})$.  Results from
different volume fractions collapse to the same universal curve.

\begin{figure}[ht!]
  \centerline{
    \includegraphics[height=0.38\textwidth]{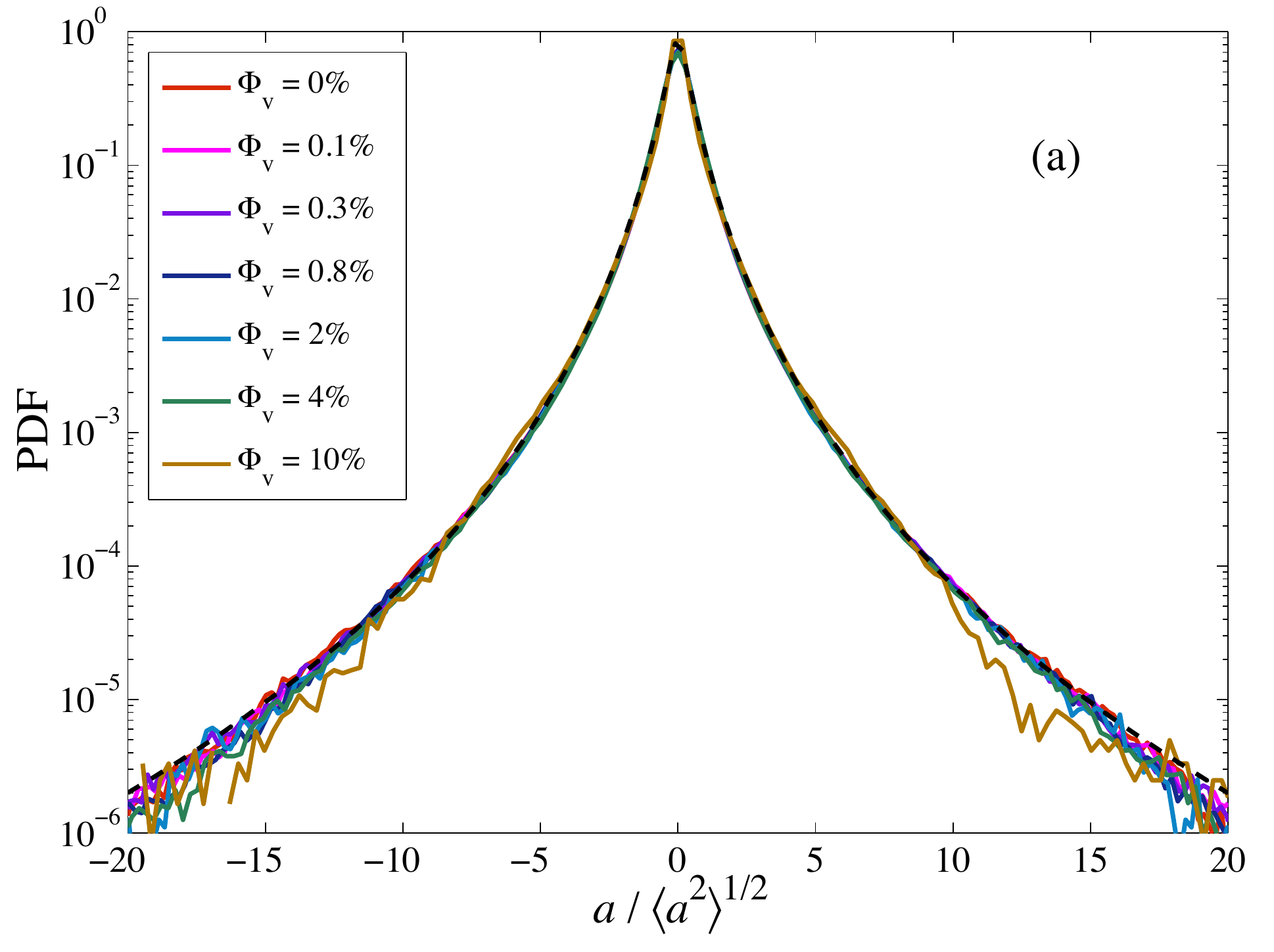}
    \hfill
    \includegraphics[height=0.38\textwidth]{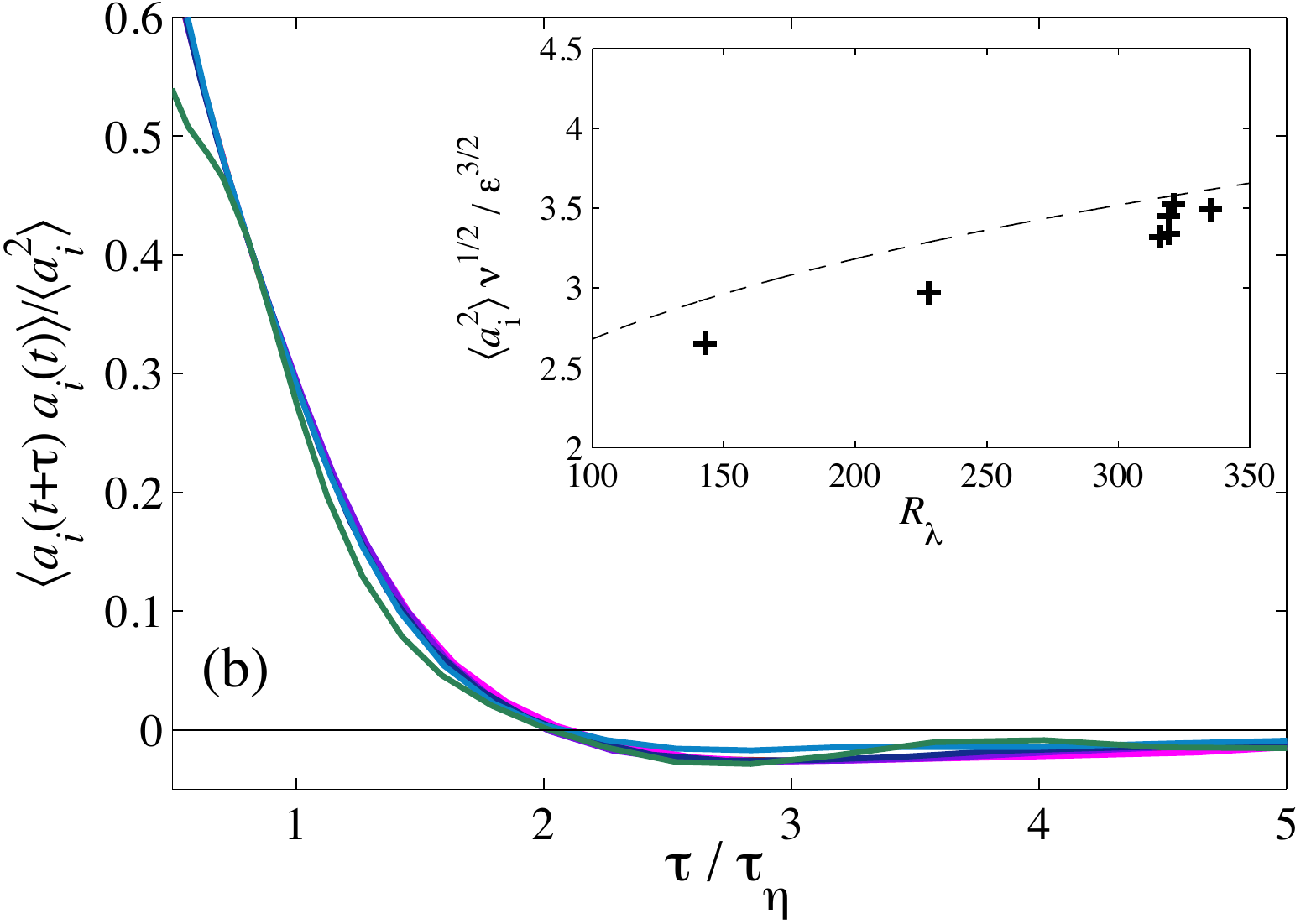}}
  \vspace{-10pt}
  \caption{\label{fig:accel} (a) Normalized probability density
    functions of acceleration components for various values of the
    volume fraction $\Phi_\mathrm{v}$ as labeled; The black dashed
    line is the approximative form obtained when assuming that the
    acceleration amplitude is log-normal.\cite{MCB04} (b) Temporal
    autocorrelation of the acceleration components for the same values
    of $\Phi_\mathrm{v}$. \emph{Inset:} Normalized variance of the
    acceleration as a function of the Reynolds number $R_\lambda$;
    Each symbol is a different experiment and the dashed line is the
    fit proposed by Hill\cite{Hill02} for the intermittent dependence
    of the acceleration variance as a function of $R_\lambda$.}
\end{figure}

The statistics of the fluid acceleration seem also not affected by the
large particles.  Figure~\ref{fig:accel}(a) shows the density
functions of the acceleration components $a_i$ after they are
normalized by their respective second-order moments.  Up to possible
slight variations in the tails that could correspond to the change in
Reynolds number at varying the fraction of large particles, the
different curves seem to collapse reasonably well.  It is known that
the moments of these distributions do not reflect the actual value of
the acceleration variance because of data filtering. To measure
$\langle a_i^2\rangle$ we follow the method of Voth \textit{et
  al.}\cite{VLC+02} consisting in extrapolating the variance when
decreasing the filter size to zero.  The resulting values of $a_0 =
\langle a_i^2\rangle \nu^{1/2}/\varepsilon^{3/2}$ are shown in the
inset of Fig.~\ref{fig:accel}(b) as a function of the Reynolds number
of the flow. One clearly observes that the decrease of $a_0$ when
$\Phi_\mathrm{v}$ increases relates to the change in Reynolds
number. Finally, the acceleration autocorrelations represented in the
main panel of Fig.~\ref{fig:accel}(b) are not much affected by the
particle presence. The slight variations are again due to the change
in Reynolds number induced by the particles.

%\bigskip 
%\noindent\emph{Possible phase transitions?}
%\smallskip

\begin{figure}[ht!]
  \centerline{
    \includegraphics[height=0.38\textwidth]{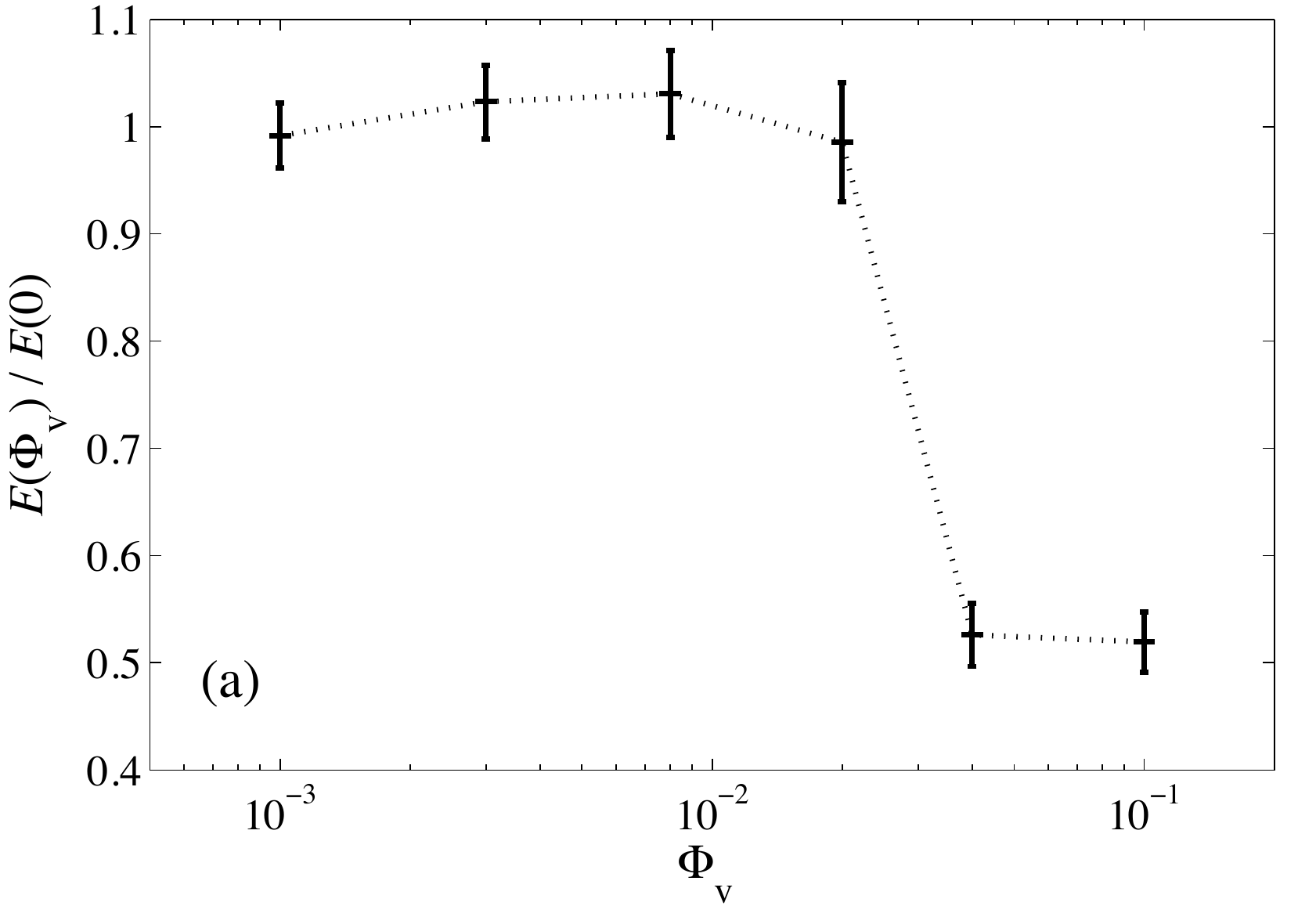}
    \hfill
    \includegraphics[height=0.42\textwidth]{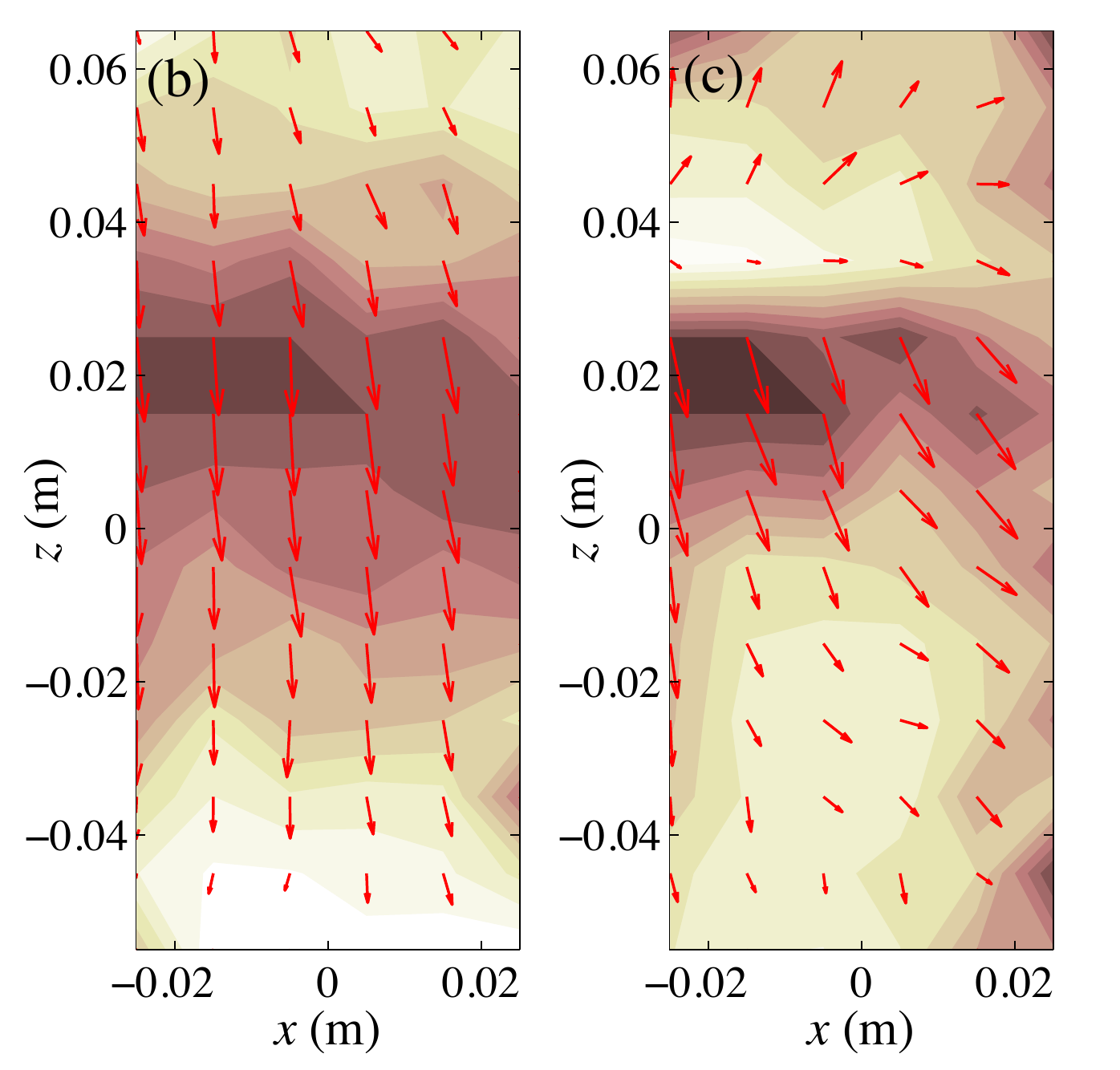}}
  \vspace{-10pt}
  \caption{\label{fig:meanflow} (a) Mean flow kinetic energy
    $E(\Phi_\mathrm{v})$ contained in the measurement volume as a
    function of the large-particle volume function. (b) and (c) Slice
    of the mean flow in a central part of the measurement volume for
    $\Phi_\mathrm{v} = 0.8\%$ and $4\%$, respectively (the $x$
    direction is along the impellers axis of rotation). The colored
    background shows the velocity squared modulus (normalized in each
    case to its mean; increasing values from white to black) and the
    arrows shows the $(x,z)$ components of the velocity field.}
\end{figure}
All the quantities considered above are either reduced as
$\Phi_\mathrm{v}^{2/3}$ or not affected by the large particles. They
relate to the turbulent fluctuations from which we have removed the
mean flow.  We actually find that the mean flow itself is strongly
affected by the large-particle load. To quantify its variations as a
function of the particle volume fraction, we have computed its kinetic
energy large-particle load. To quantify its variations as a function
of the particle volume fraction, we have computed its kinetic energy
$E(\Phi_\mathrm{v}) = \int |\langle \bm u(\bm x) \rangle|^2
\mathrm{d}^3x$.  Figure~\ref{fig:meanflow}(a) shows that
$E(\Phi_\mathrm{v})$ remains almost constant up to $\approx 3\%$ where
it suddenly drops down. Above this critical value, the mean flow is in
a different state that contains approximately twice less energy. The
transition occurs roughly when the typical distance between particles
becomes less than their diameter. All particles are thus necessarily
feeling the presence of their neighbors through the flow
modifications.  This apparently has a drastic impact on the average
motions as can be seen when comparing Figs.~\ref{fig:meanflow}(b) and
(c). They represents the variations of the energy and of the $(x,z)$
components of the mean flow in the same slice of the measurement
volume for $\Phi_\mathrm{v} = 0.8\%$ (b) and $4\%$ (c). Not only is
the energy content strongly modified but also the fine structures and
the directions of the mean flow. Panel (b) is representative of all
volume fractions below $0.8\%$ while panel (c) is also representative
of $\Phi_\mathrm{v} = 10\%$. For $\Phi_\mathrm{v} = 2\%$, the mean
flow obtained when time-averaging is very different from these two
states. We however suspect it to be non stationary because several
movies show extremely strong deviations from the mean. Unfortunately,
our experiment was conceived to focus on small-scale turbulent
properties and not designed to accurately measure large scale
variations.  A better understanding of what is happening for this
volume fraction would require revisiting our setup in order to focus
on the large scales. This would also allow us a more accurate study of
a possible phase transition for $\Phi_\mathrm{v}$ of the order of a
few percents.

To conclude, let us stress again our main findings on the modification
of a turbulent von K\'arm\'an flow by finite-size neutrally-buoyant
spherical particles.  The most surprising result is that
inertial-range and small-scale turbulent features are unchanged in the
presence of particles for volume fractions up to
$\Phi_\mathrm{v}=10\%$. There is no signature of the particle size and
its associated timescale neither on the scaling properties of the
fluid turbulent velocity fluctuations nor on the statistics of the
acceleration.  The only noticeable effects are on global quantities.
In our specific setup and when the impeller rotation rate is
maintained, the turbulent kinetic energy and the inertial-range
transfer rate of the bulk flow are continuously decreasing when the
particle volume fraction is increased.  However, our setup does not
monitor the energy injection rate.  We are thus unable to discriminate
between a decrease of the power needed to maintain the impellers
rotation speed (i.e.\ a sort of drag reduction) and a redistribution
of the energy dissipation to the regions closer to the impellers
(outside of our observation volume).  A better understanding would
require a more accurate handling of the impeller torque and thorough
measurements of the fluid velocity in the full experimental domain.
Such measures could clarify the possible phase transition at
$\Phi_\mathrm{v}\approx3\%$.  In particular, they would allow us to
draw differences between what is due to the specificities of von
K\'arm\'an inhomogeneous flow and a possible universal change at such
volume fractions in the coupling between the turbulent flow and the
particles.

\smallskip
This research has received funding from the European Research Council
under the European Community's Seventh Framework Program
(FP7/2007-2013 Grant Agreement No.~240579) and from the French Agence
Nationale de la Recherche (Programme Blanc ANR-12-BS09-011-04).

\bibliography{biblio}

\end{document}